# Atomic size zone interaction potential between two ground-state cold atoms

Zhaoying Wang[1,*], Xuehan Jin[1], Yunhan Wu[1]

[1]Institute of Optics, Department of Physics, Zhejiang University, Hangzhou 310027, China

Corresponding authors: *zhaoyingwang@zju.edu.cn

The complex-source-point model are already used in the exact solution for the urtrashort pulse and nonparaxial beam. In this letter we have used the complex-source-point model to deduce the interaction potential equation for the separation $R$ between two atoms which is comparable with the size of the atoms. We show the result and the characteristics of the numerical calculation. Since the singular point around $R$=0 is removed by using the complex-source-point model, so that we can obtain the result force around $R$=0. With the decreasing of the distance between two atoms, the force switches from the electromagnetic force to the strong force by use our equation.

## I. INTRODUCTION

In 1948, the famous Dutch physicist H.B.G. Casimir predicted the existence of a force between two conducting plates [1]. The same year, Casimir together with Polder predicted the analogous attractive force between two atoms (or molecules) [2]. The Casimir-Polder (CP) force arises from quantum fluctuations of the electromagnetic field. For atoms having a dominant transition with frequency $\omega_0$ between the ground and first excited states, they showed that the interaction between the two atoms reduced to the London limit of the van der Waals interaction in the near zone, i.e., $R^{-6}$ dependence for small separations ($R\omega_0 \ll 1$). In contrast, the interaction energy decays like $R^{-7}$ in the far zone [2].

On the experimental side, CP forces measurements at distances ranging from $10^{-8}$ m to $10^{-3}$ m have been the domain of microelectromechanical systems (MEMS) and of torsion balance experiments [3-5]. Over the last decade, a number of proposals and experiments that explore CP interactions at very short range using the metrological advantages of cold atoms and Bose Einstein condensates (BEC). For instance, atoms reflecting from evanescent waves near surfaces [6], atoms transmitted through a cavity made of two gold coated plates [7], atoms diffracted from a material grating [8], atoms undergoing quantum reflection [9], the frequency shift of the collective oscillations of a magnetically trapped BEC [10, 11], BEC vibrations by using micromechanical cantilever [12], Bloch oscillations [13], atom interference phase difference [14] and atomic clock shift [15], etc. Since the cold atoms has less moving velocity and even can be almost stopped at one position, so that the separation between two atoms can be adjusted much shorter than the traditional mechanical system. A natural question along that line is what will happen when the separation between two atoms is comparable with the size of the atoms. This what we are going to investigate the atomic size zone Casimir-Polder potential between two ground-state atoms.

There are lots of methods aimed at obtaining the CP potential, for example, the perturbation theory based on two-transverse-photon exchange [16,17], the use of the multipolar framework include response theory [18], consideration of the changes in zero-point energy[19], radiative reaction[20], evaluation of energy shifts in the Heisenberg picture[21], the method based on spatial vacuum field correlations[22-24], source theory[18] and so on. For the calculation of van der Waals forces between nanoparticles whose size and separation are comparable, traditional procedures via summing over infinitesimal elements, such as Hamaker's method[25] and the proximity force approximation[26], have been widely used. However, van der Waals forces are inherently nonadditive interactions. Recently, Zhao et al. study the van der Waals forces between two spheres and between a sphere and a plane with arbitrary separation using transformation optics [27].

Our calculation of the interaction potential between two atoms is based upon the method of equal-time spatial vacuum field correlations [22-24], which can simplify some calculations in some complex external environment. The main idea based on the vacuum spatial correlations can be explained as that the vacuum fluctuations of the electromagnetic field induce instantaneous correlated dipole moments on the two atoms, and the interaction potential energy can be obtained by calculating the classical interaction between the two correlated induced dipoles [22-24].

## II. THE ELECTROMAGNETIC FIELD OF THE INDUCED DIPOLE

Let us assume a pair of electric dipoles in vacuum. The negative charge $+q$ is fixed at the origin of the coordinates, while the positive charge $-q$ The electromagnetic field of the induced dipole is located along the $x$ axis and is oscillating with time. Thus the time-varying electric dipole moment can be written as $\vec{p}(t) = ql(t)\vec{e}_x$, where $l(t)$ is the distance between the negative and positive charges and $\vec{e}_x$ is the unit vector along the $x$ axis. Under the Lorentz condition, using the relations between the electromagnetic field and the potential function, we can derive the expression for the electric dipole radiation field as the following [28]:

$$\vec{E}(x,y,z,t) = -\frac{c^2 \mu_0}{4\pi} \left\{ \frac{[\ddot{p}]}{c^2 R} + \frac{[\dot{p}]}{cR^2} + \frac{[p]}{R^3} \right\} \vec{e}_x + \frac{c^2 \mu_0 x}{4\pi R^2} \\ \times \left\{ \frac{[\ddot{p}]}{c^2 R} + \frac{3[\dot{p}]}{cR^2} + \frac{3[p]}{R^3} \right\} (x\vec{e}_x + y\vec{e}_y + z\vec{e}_z), \quad (1)$$

$$\vec{H}(x,y,z,t) = -\frac{c}{4\pi R} \left\{ \frac{[\ddot{p}]}{c^2 R} + \frac{[\dot{p}]}{cR^2} \right\} (z\vec{e}_y - y\vec{e}_z) \quad (2)$$

Where $R = \sqrt{x^2 + y^2 + z^2}$ is the distance from the origin to the observation point, $c$ is the velocity of light in vacuum, the time factor of the physical quantities written in shortened form with brackets indicates the retarded time ($t$-$R/c$), $[\dot{p}]$ and $[\ddot{p}]$ represent the first and second order partial derivatives of the electrical dipole moment p with respect to the retarded time respectively.

In order to compare the expression with the others [22-24], we modify the electric field expression as the following:

$$E_i(\omega_k, R) = \frac{c^2 \mu_0}{4\pi} D_{ij}^R(\omega_k, R)\left[p_j(\omega_k)\right],\qquad(3)$$

$$D_{ij}^R(\omega_k, R) = e^{ikR}\left[\left(\frac{k^2}{R} + \frac{ik}{R^2} - \frac{1}{R^3}\right)(\delta_{ij} - \hat{R}_i\hat{R}_j) + 2\left(\frac{1}{R^3} - \frac{ik}{R^2}\right)\hat{R}_i\hat{R}_j\right].\qquad(4)$$

Where $\omega_k$ is the oscillating frequency of the dipole, and $k = \omega_k/c$ is the wave number. $\hat{R}_i = R_i/R$ denotes the *i*th element of the unit displacement vector of $\vec{R}/R$.

Assuming there are two atoms A and B fixed at the certain locations in space-time. The distance between two atom's center is $R = 2r_0 + \Delta$, where $r_0$ is the radius of the atom and $\Delta$ is the separation of two atom's edge. The potential energy then can be considered as the interaction between the two correlated induced dipoles according to the spatially correlated vacuum fluctuations. Therefore, the interaction potential of two ground-state atoms reads [22-24]

$$V_{AB}(R) = -\hbar\,\mathrm{Im}\int_{-\infty}^{+\infty}\frac{d\omega_k}{2\pi}\coth\frac{\hbar\omega_k}{2kT}\left[\alpha_A(\omega_k)\alpha_B(\omega_k)D_{ij}^R(\omega_k,R)D_{ij}^R(\omega_k,R)\right]\qquad(5)$$

Where the term $\coth(\hbar\omega_k/2kT)$ is called as the temperature factor. $\alpha_{A,B}(\omega_k)$ is the isotropic polarizability of atoms. For the case of a two-level atom, the isotropic polarizability can be written as [24,29]

$$\alpha(\omega_k) = \frac{2\omega_0 \mu^2}{3(\omega_0^2 - \omega_k^2)}\qquad(6)$$

Where $\omega_0$ is the transition frequency from the ground sate to the excited state and $\mu$ is the matrix element of the atomic dipole momentum operator. There are one special example needed to discuss, for cold atoms $kT \to 0$, then $\coth(\hbar\omega_k/2kT) \to 1$. Substituting Eq.(4) into Eq.(5), the interaction potential of two cold atoms becomes

$$V_{AB}(R) = -\frac{\hbar c}{\pi R^2} \mathrm{Im} \int_o^{+\infty} \alpha_A(\omega_k) \alpha_B(\omega_k) k^4 e^{2ikR} \left[1 + \frac{2i}{kR} - \frac{5}{k^2 R^2} - \frac{6i}{k^3 R^3} + \frac{3}{k^4 R^4}\right] dk \quad (7)$$

Considering that the molecular polarizability is real at real and imaginary frequency, the potential can be written in the more familiar form in terms of an integral over imaginary frequencies [23].

$$V_{AB}(R) = -\frac{\hbar c}{\pi R^2} \int_o^{\infty} \alpha_A(iu) \alpha_B(iu) u^4 e^{-2uR} \left[1 + \frac{2}{uR} + \frac{5}{u^2 R^2} + \frac{6}{u^3 R^3} + \frac{3}{u^4 R^4}\right] du \quad (8)$$

where $u = -ik$. We know that there is a singular point at $R=0$ in Eq. (8). Usually, the distance between atoms $R$ is much larger than the atom size. But with the lower of the atom temperature, the distance between atoms can be comparable with the atom size. So it is important to analyze the interaction potential of two atoms for $R$ close to zero.

## III. THE NUMERICAL CALCULATION OF THE INTERACTION POTENTIAL

In order to remove the singular point, we use the complex-source-point model, which was first introduced for scalar beams by Deschamps[30]. The basic idea of the complex-source-point (CSP) model is as follows. As shown in Fig.1, every atom is composed of the nucleus and the electron cloud. Electrons surround the atomic nucleus in pathways called orbitals. The inner orbitals surrounding the atom are spherical but the outer orbitals are much more complicated. Due to the position of the electron is not fixed, so that the distance between two atoms depends on the position of dipoles, which is always changed. Thus we can consider the distance between two atoms as $R' = R + iR_0$ according to the complex-source-point model. $R$ is the distance between two centers of the atom nucleus. $R_0$ is the coordinate value of the source point on the imaginary axis. The value of the $R_0$ is related to the atom size, and explain the uncertainty of the position of the electron. By the way, in the complex-source-point model, the electric dipole radiation fields still satisfy Maxwell's equation exactly [28].

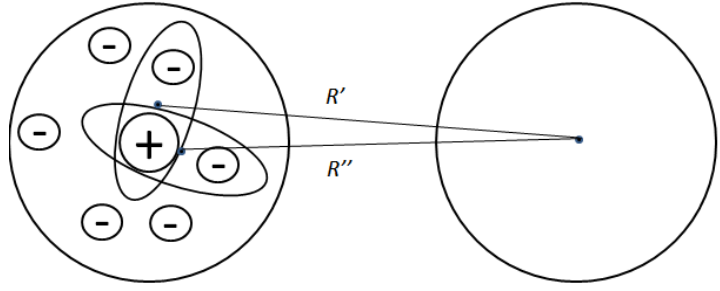

Fig.1 the basic complex-source-point (CSP) model for interaction potential between two atoms.

Substituting $R'$ into Eq.(8), considering that the complex-source-point model has clear physical significance only in the real space, only the real part of the potential is needed. Now we illustrate the interaction potential in the complex-source-point model with some numerical calculation. Assume $k_0=2\pi/780 nm^{-1}$, $\mu_0=1$ and $R_0=10^{-11}$m. Of course, the parameters can be other value, which does not make essential difference in the result. Fig.2(a)-(c) show the interaction potential versus the $R$ coordinate in the complex-source-point model. Generally, the interaction potential approaches to infinite as $R$ is close to zero because of the singular point at $R=0$, as shown in Fig. 2(d). But the interaction potential in the complex-source-point model is different, there is three singular value of the potential nearby $R=0$. The position of these three singular points are around at $0.5R_0$, $1.2R_0$, $4.2R_0$ respectively, the three coefficients depend on the oscillating frequency of the dipole. Here, we should explain the calculation of the different value of $R_0$. The value of the interaction potential in the complex-source-point model will be decreased with the increasing of the value of $R_0$, but the three singular points are still lie around at $0.5R_0$, $1.2R_0$, $4.2R_0$ respectively.

From Fig. 2(a), we can find that when the distance $R$ is smaller than the first singular value around $0.5R_0$, the interaction force is strong repulsive force. With the distance increasing, the interaction force become attractive force. When the distance $R$ is bigger than 1.2 $R_0$, the interaction between two atoms is repulsive force, until $R$ colse to 4.2 $R_0$, then force become attractive force again, as shown in Fig. 2(b). This behavior like the van der Waals force. After the distance increasing to $20R_0$, the value of interaction force is almost the same as the general CP force, as shown in Fig.2(c) and (d).

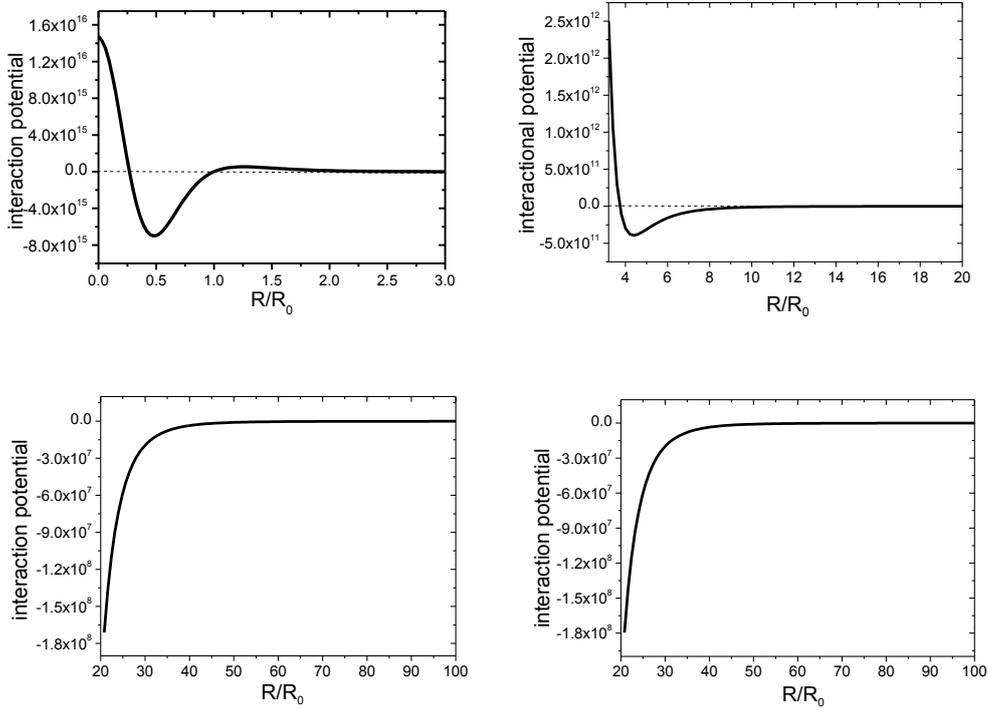

FIG. 2. (a)-(c) The interaction potential between two atoms in the complex-source-point model with $R_0 =10^{-11}$. (d) The general CP potential.

As we shown above, when the distance bigger than $1.2R_0$, the interaction force behavior as the van der Waals force and CP force, which are the electromagnetic interaction force. Now, we focus on the interaction potential for the distance smaller than $1.2R_0$. We think this interaction belongs to strong interaction, include the strong attractive force and the strong pulsive force. The value of strong force is about $10^4$ bigger than the electromagnetic force. When the distance is between $0.5R_0$ and $1.2R_0$, the strong interaction force is attractive, which force overcomes the repulsive van der

Waals force, make the hadrons bound together tightly into the atom nuclear. When the distance between the two atoms is smaller than $0.5R_0$, the strong interaction force is pulsive, which force prevents the two atoms from overlapping together.

## IV. CONCLUSION

In this letter we have used the complex-source-point model to study the interaction potential in the separation between two atoms which is comparable with the size of the atoms. We show the result and the characteristics of the numerical calculation. With the increasing of the distance between two atoms, there are four kinds of force, which corresponding to two strong forces (replusive and attractive) and two electromagnetic forces (replusive and attractive). The electromagnetic force is the same behavior like the van der Waals force. So that in this paper, we would like to say that we might already use one equation to unify two forces: strong force and electromagnetic force.

## ACKNOWLEDGMENTS

The authors wish to acknowledge financial support from the National Natural Science Foundation of China under Grant No 11174249, the National High-Technology Research and Development of China under Grant No 2011AA060504, and the Fundamental Research Funds for the Central Universities No. 2016FZA3004.